\documentclass[%
 aip,
 amsmath,amssymb,
 reprint,%
]{revtex4-1}

\usepackage{graphicx}
\usepackage{dcolumn}
\usepackage{bm}

\usepackage[utf8]{inputenc}
\usepackage[T1]{fontenc}
\usepackage{mathptmx}
\usepackage{etoolbox}

\usepackage[table]{xcolor}
\usepackage{array}
\usepackage{tabularx}
\usepackage{booktabs}
\usepackage{makecell}
\usepackage{ragged2e}
\usepackage{threeparttable}

\definecolor{goodgreen}{RGB}{225,242,215}
\definecolor{badred}{RGB}{252,225,225}
\definecolor{unknowngray}{RGB}{242,242,242}
\definecolor{headerblue}{RGB}{225,233,246}

\newcolumntype{L}[1]{
  >{\RaggedRight\arraybackslash}p{#1}
}

\newcolumntype{Y}{
  >{\RaggedRight\arraybackslash}X
}

\makeatletter
\def\@email#1#2{%
 \endgroup
 \patchcmd{\titleblock@produce}
  {\frontmatter@RRAPformat}
  {\frontmatter@RRAPformat{\produce@RRAP{*#1\href{mailto:#2}{#2}}}\frontmatter@RRAPformat}
  {}{}
}%
\makeatother
\begin{document}

\preprint{AIP/123-QED}

\title{Highly Scalable Selectorless Cryogenic Memory Array Using Ferroelectric Josephson Field-Effect Transistors}
\author{Saheeb Ahmad}
\affiliation{Holcombe Department of Electrical and Computer Engineering,
Clemson University, Clemson, South Carolina 29634, USA}

\author{Shamiul Alam}
\email{shamiua@clemson.edu}
\affiliation{Holcombe Department of Electrical and Computer Engineering,
Clemson University, Clemson, South Carolina 29634, USA}

\begin{abstract}
Scalable memory systems that satisfy the temperature, speed, and energy requirements of cryogenic environments are essential for the development of large-scale quantum computers. They may also benefit high-performance computing and space applications. However, existing cryogenic memory technologies often suffer from limited scalability, low operating speed, and/or high power consumption, restricting the scalability of target applications.
Ferroelectric Josephson field-effect transistors (Fe--JoFETs), which combine ferroelectric polarization with the superconducting properties of Josephson junctions, offer a promising solution. The ferroelectric layer enables nonvolatile storage capability, while the Josephson junction supports high-speed, energy-efficient operations. In this work, we leverage Fe--JoFETs to develop a highly scalable, ultra-low-power, nonvolatile cryogenic memory array that does not need additional selector devices for random access. Moreover, the superconducting component of Fe--JoFET provides a binary decision during read, eliminating the need for sensing peripheral circuitry. We first develop a physics-based Verilog-A compact model for Fe--JoFETs and use it to verify the functionality of the proposed memory array. By eliminating both selector and sensing circuitry, the proposed memory architecture offers higher scalability than existing technologies. The ultra-low-power operation of this memory also makes it compatible with strict power budgets of cryogenic applications.
\end{abstract}

\maketitle

\section{\label{sec:level1}Introduction}

To develop a quantum computer with thousands of qubits to realize its full potential, we need cryogenic control processors and memory systems. In existing quantum computers, superconducting (SC) qubits are placed at tens of millikelvin temperature, while the control
processor and memory remain at room temperature \cite{Alam2023CryogenicMemory,Hornibrook2015CryogenicControl}.
Placing the memory far from the SC qubits increases the
number of electrical connections and leads to greater thermal leakage,
latency, and wiring complexity
\cite{Alam2023CryogenicMemory,Hornibrook2015CryogenicControl}, eventually limiting the number of qubits. Large
quantum algorithms also require substantial memory for storing program instructions, calibration data, intermediate results, and continuous error-correction operations \cite{Tannu2017CryoDRAM,Alam2023CryogenicMemory}.
In addition, a suitable and scalable cryogenic memory can be beneficial for high-performance computing and space electronics, where low-power and high-speed operations are required within strict cooling budgets
\cite{Likharev1991RSFQ,Holmes2013EnergyBudget}.

Existing cryogenic memory technologies include (i) non-superconducting memories \cite{Tannu2017CryoDRAM, Shu2023CSDBeDRAM} such as SRAM, DRAM, resistive, spintronic, and ferroelectric, (ii) SC memories such as Josephson junction (JJ)-based \cite{Kirichenko1999}, magnetic JJ-based \cite{Ryazanov2012MJJMemory}, superconducting memristor-based \cite{Alam2021CryoMemristor}, ferroelectric SC quantum interference device (FeSQUID)-based \cite{Alam2022FeSQUIDMemory}, and cryotron-based memories, and (iii) hybrid memories such as Josephson-CMOS, cryotron-MTJ, and Josephson-spintronic memories \cite{Alam2023CryogenicMemory, alam2021non}. Non-superconducting memories offer technological maturity and relatively high storage
capacity, but their speed, power consumption, operating temperature, and
interface requirements are incompatible with SC
processors operating near 4 K
\cite{Alam2023CryogenicMemory,Tannu2017CryoDRAM}. On the other hand, SC memories (based on JJs and SQUIDs) offer the most favorable combination of energy efficiency and operating speed \cite{Anderson1963JosephsonObservation,Josephson1962Tunnelling,Jaklevic1964SQUID}. However, conventional implementations often require large cells, inductive coupling, transformers, and auxiliary junctions for addressing, making these systems suffer from poor scalability \cite{Alam2023CryogenicMemory,Tolpygo2016Scalability,Holmes2013EnergyBudget, Alam2023FeSQUIDMVM}.
Emerging solutions based on superconducting memristors, cryotrons, and FeSQUIDs can combine the density of non-SC memories
with the speed of SC circuits, but they require additional access devices to allow random access, among other issues.
\cite{Alam2022FeSQUIDMemory,Alam2021CryoMemristor, Alam2023CryogenicMemory}.

FeSQUID devices that use the interplay between ferroelectric materials and SC devices, have been utilized to develop cryogenic memories, voltage-controlled logic circuits, and in-memory computing systems \cite{Alam2025FerroelectricSuperconducting, Alam2023FeSQUIDMVM, Alam2024FeSQUIDLogic, Alam2022FeSQUIDMemory, alam2025tcam}. However, FeSQUID-based memory suffers from the requirement of an access device (heater cryotron) which imposes additional limits on the scalability, speed, and power consumption of the memory system \cite{Alam2022FeSQUIDMemory}.
Recently, a new type of ferroelectric-controlled SC device has been experimentally demonstrated called a ferroelectric Josephson field-effect transistor (Fe--JoFET) \cite{Paghi2025FeJoFET}. Due to its ferroelectric-based gate, the device exhibits voltage-controlled polarization switching, which eventually affects the critical current ($I_c$) and normal-state resistance ($R_N$) of its SC channel \cite{Paghi2025FeJoFET}. This allows the device to operate in resistive and SC states, which can be controlled through gate voltage \cite{Paghi2025FeJoFET}. As such, the Josephson channel provides high-speed and energy-efficient transport, while the ferroelectric gate provides direct voltage control and nonvolatile
storage. This also enables the stored state to be identified directly from the SC ($0V$) or
resistive ($\neq 0V$) response of the selected device, eliminating the need for additional sensing circuitry.

\begin{figure*}[t]
    \centering
    \includegraphics[width=0.75\textwidth]{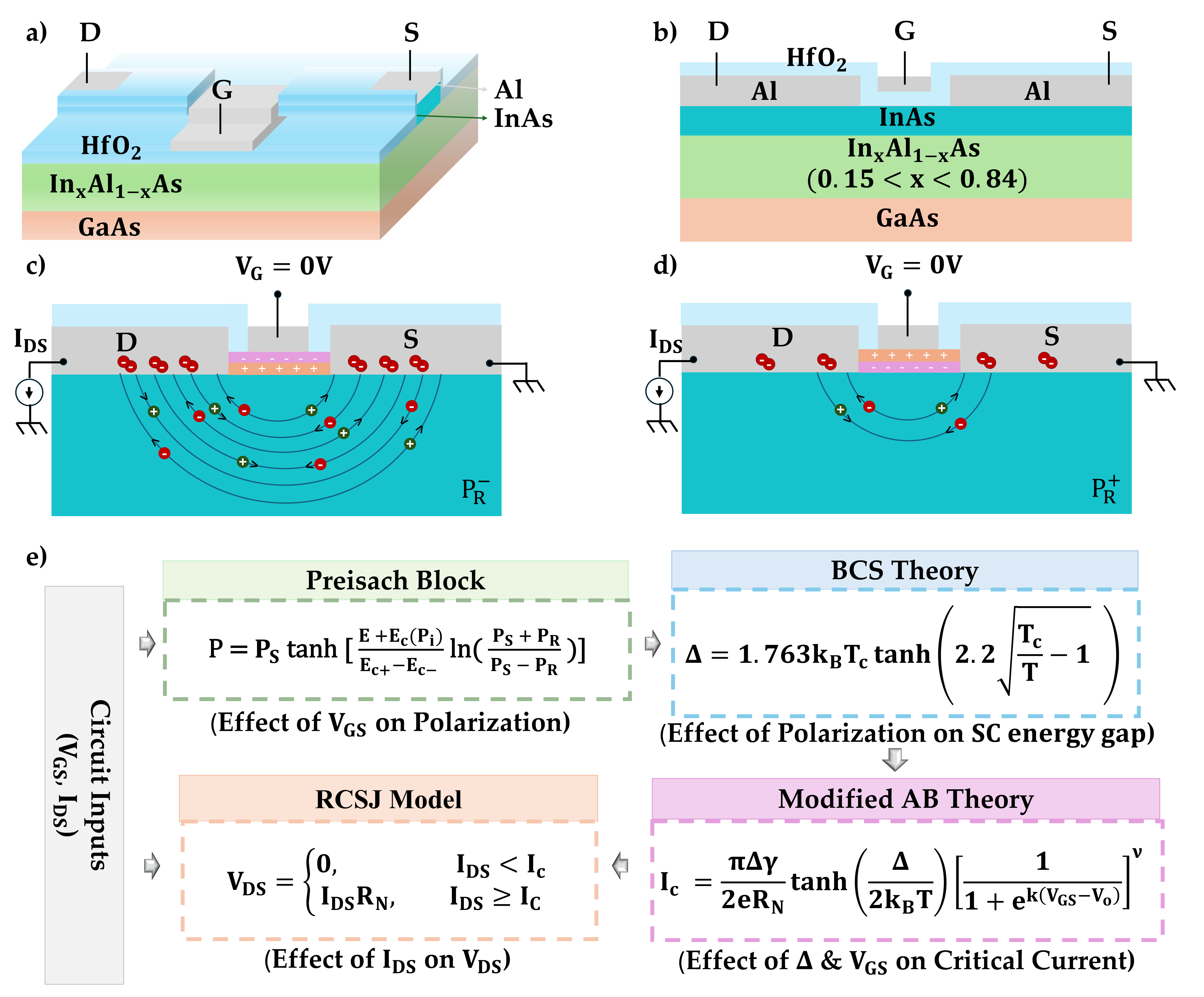}
    \caption{Device structure and compact modeling methodology of Fe--JoFET devices. \textbf{(a)} Device structure of the Fe--JoFET. \textbf{(b)} Cross-sectional view of the device stack. Carrier distribution in the $\mathrm{InAs}$ channel for \textbf{(c)} negative ($P_{\mathrm{R}}^{-}$) and \textbf{(d)} positive ($P_{\mathrm{R}}^{+}$) remnant-polarization states at $V_{\mathrm{GS}}=0$ V. The two remnant-polarization states are achieved by applying suitable voltage across the ferroelectric layer, which modulates the carrier density in the InAs channel, thereby controlling the SC to resistive switching of the Fe--JoFET channel. \textbf{(e)} Block-level representation of the developed Verilog-A compact model.}
    \label{fig:FeJoFET_Device_Model}
\end{figure*} 

In this work, we design a nonvolatile, selectorless, and energy-efficient cryogenic memory array using Fe--JoFETs that can solve the existing challenges and facilitate the development of a suitable and scalable cryogenic memory system. During write operations, all cells remain superconducting, while during read, only the
accessed cell may be in the resistive state (for logic '1'). Therefore, the array minimizes voltage drops and resistive dissipation as well as reducing the risk of disturbing the unaccessed cells.
The resulting combination of nonvolatile storage, intrinsic binary
state discrimination, selectorless access, and SC operation
of unaccessed cells provides a highly energy-efficient and scalable
approach to cryogenic memory systems. Although Fe--JoFET has been experimentally demonstrated, to the best of our knowledge, no compact model is available. Therefore, to explore its potential at the circuit and system-level and to verify the functionality of our memory design, we initially develop a physics-based compact model for Fe--JoFETs. The developed model uses (i) a Preisach block \cite{Preisach1935,Ni2018FeFETModel} for polarization switching of ferroelectric materials, (ii) a modified Bardeen--Cooper--Schrieffer (BCS) theory \cite{Bardeen1957BCS} for polarization-dependent SC energy gap ($\Delta$), (iii) a modified Ambegaokar--Baratoff (AB) theory\cite {Ambegaokar1963Tunneling} for gate voltage and polarization-dependent $I_c$, and (iv) the resistively and capacitively shunted junction (RCSJ) model for current-voltage characteristics  
\cite{Stewart1968JosephsonIV,McCumber1968WeakLink}.  

\begin{figure*}[t]
    \centering
    \includegraphics[width=0.85\textwidth]{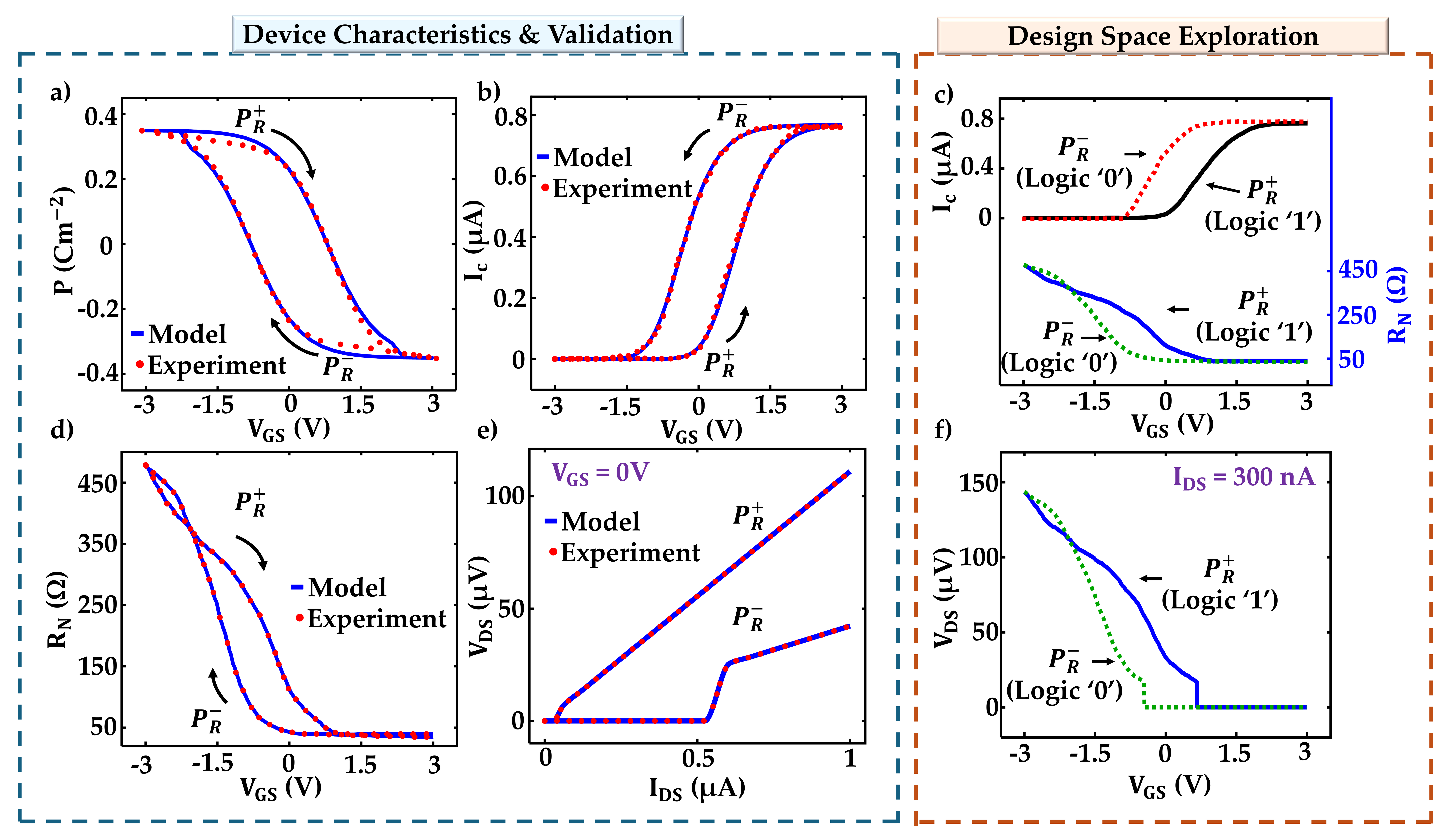}
    \caption{Device characteristics, model validation, and design space exploration of Fe--JoFET. \textbf{(a)} $P$ vs. $V_{\mathrm{GS}}$, \textbf{(b)} $I_{\mathrm{c}}$ vs. $V_{\mathrm{GS}}$, \textbf{(d)} $R_{\mathrm{N}}$ vs. $V_{\mathrm{GS}}$, and \textbf{(e)} $V_{\mathrm{DS}}$ vs. $I_{\mathrm{DS}}$ characteristics for Fe--JoFETs.  In panels (a), (b), (d), and (e), the blue curves and red dots represent the proposed compact-model results and experimental data, respectively, demonstrating the ability of the model to reproduce the device characteristics of the Fe--JoFET. Model-predicted \textbf{(c)} $I_{\mathrm{c}}$ and $R_{\mathrm{N}}$ vs. $V_{\mathrm{GS}}$ \textbf{(f)} $V_{\mathrm{DS}}$ vs. $V_{\mathrm{GS}}$, for $P_{\mathrm{R}}^{+}$ and $P_{\mathrm{R}}^{-}$ states. The experimental data used for validation are obtained from Ref. \cite{Paghi2025FeJoFET} using a Fe--JoFET device with a width of $5~\mu{m}$ and a length of $600~{nm}$ at an operating temperature of $50~{mK}$.}
    \label{fig:FeJoFET_Characteristics_Exploration}
\end{figure*}

\section{Fe--JoFET and Device Characteristics}

Fe--JoFET is a Josephson FET on an InAs on insulator platform as represented in Fig.~\ref{fig:FeJoFET_Device_Model}(a) and (b). The insulating layer consists of GaAs with an In$_x$Al$_{1-x}$As metamorphic buffer and an intrinsically n-type InAs epilayer. Finally, the gate stack comprises an HfO$_2$ ferroelectric gate layer and Al SC layers for the source and drain leads. Detailed descriptions of the device fabrication process, its key dimensions, and experimental parameters are reported in Ref. \cite{Paghi2025FeJoFET}.

The ferroelectric hysteresis of the $\mathrm{HfO}_2$ layer produces two
distinct polarization states through its remnant polarization ($P_R$). $P_R$ can be controlled via an applied electric field, $E$ ($=-V_{GS}/d$, where $V_{GS}$ and $d$ are the gate-to-source voltage and the ferroelectric thickness, respectively). The application of a sufficiently positive $E$ (or negative $V_{GS}$) and a sufficiently negative $E$ (or positive $V_{GS}$) results in the positive ($P_R^+$) and negative ($P_R^-$) remnant polarization states, respectively. These polarization states then affect the modulation of charge carriers in the proximitized $\mathrm{InAs}$ epilayer. The $P_R^-$ state results in an enhancement of electrons in the $\mathrm{InAs}$ semiconducting layer, whereas the $P_R^+$ state results in an activation of charge trapping, which causes a depletion of electrons from the $\mathrm{InAs}$ epilayer into interfacial traps, as shown in Fig.~\ref{fig:FeJoFET_Device_Model}(c) and (d). Now, higher number of majority carriers leads to higher $I_c$ and lower $R_N$. Therefore, the $P_R^-$ state of the ferroelectric causes a higher $I_c$ and lower $R_N$ than that of the $P_R^+$ state. 

Thus, the ferroelectric $P_R$ states of the Fe--JoFET allow the device to operate in two distinct nonvolatile states, enabling binary information storage. Fe--JoFETs show prominent signs of ferroelectric $P$ for a $V_{GS}$ scan of [-2,2] V, with ranges below that level introducing negligible hysteresis. For our work, we utilize the experimental data reported in Ref.\cite{Paghi2025FeJoFET} for $V_{GS}$ scans of [-3,3] V.

\section{Modeling Methodology, Validation, and Design Space Exploration}
\subsection{Methodology}\label{Methodology}

Our compact model follows the methodology shown in Fig.~\ref{fig:FeJoFET_Device_Model}(e). 
Ferroelectric materials exhibit nonvolatile polarization states controlled by voltage bias. Preisach models \cite{Preisach1935,Ni2018FeFETModel} are widely used for their ability to accurately represent this polarization-voltage characteristics. Therefore, we model the device's $P$ using the following equation-
\begin{equation}
P = P_S \tanh\!\left[
\frac{E+E_c(P_i)}{E_{c+}-E_{c-}}
\ln\!\left(\frac{P_S+P_R}{P_S-P_R}\right)
\right]
 \label{eqn:p-v}
\end{equation} 
where $P_S$, $P_R$, and $P_i$ are the saturation, remnant, and initial polarization values. $P_R$ of any ferroelectric material can be positive ($P_R^+$) or negative ($P_R^-$) which are described by $P_i$. $E_c$ is the coercive electric field, which is either positive ($E_{c+}$) or negative ($E_{c-}$) depending on $P_i$. $E$ represents the electric field across the ferroelectric layer. We extract the values of different parameters used in Eq.~\ref{eqn:p-v} from experimental data reported in Ref. \cite{Paghi2025FeJoFET} and calibrate the model.

The two polarization states are controlled by $V_{GS}$: a large negative $V_{GS}$ (or, positive $E$) for the $P_R^+$ state and a large positive $V_{GS}$ (or, negative $E$) for the $P_R^-$ state. The positive (negative) $P_R$ states decrease (increase) the majority-carrier density in the SC channel, resulting in a change in critical temperature ($T_c$), which in turn impacts $\Delta$. Thus, the two polarization states offer two distinct values of $\Delta$, which are calculated using the BCS theory \cite{Bardeen1957BCS}-
\begin{equation}
\Delta = 1.763\,k_B T_c \tanh\!\left(2.2\sqrt{\frac{T_c}{T}-1}\right)
\end{equation}
where $k_B$ is the Boltzmann constant and $T$ is the temperature. $\Delta$, determined from ferroelectric polarization, affects $I_c$ of the SC channel, which is typically calculated using the AB theory \cite{Ambegaokar1963Tunneling}. However, to capture the effect of $V_{GS}$, we modify the AB equation. The modified AB equation is as follows-
\begin{equation}
I_c = \frac{\pi \Delta \gamma}{2 e R_N}
\tanh\!\left(\frac{\Delta}{2 k_B T}\right)
\left[\frac{1}{1 + e^{k\left(V_{GS}-V_o\right)}}\right]^{\nu}
\end{equation}
where $V_o$ is the gate-voltage offset, and $\gamma$, $k$, and $\nu$ are the fitting parameters used to modify the AB equation to capture the dependence of $I_c$ on $V_{GS}$.

The channel $I$--$V$ characteristics is derived from the RCSJ model. If the channel current ($I_{DS}$) is below (above) $I_c$, it shows zero (non-zero) resistance \cite{Stewart1968JosephsonIV,Tinkham1996Superconductivity,alam2020compact}. We model this as-
\begin{equation}
V_{DS} =
\begin{cases}
0, & I_{DS} < I_c \\
I_{DS}R_N, & I_{DS} \ge I_c
\end{cases}
\end{equation}
where $V_{DS}$ is the drain--source voltage. In Fe--JoFETs, $P_R^-$ and $P_R^+$ states will result in a high ($I_{c,\mathrm{'0'}}$) and low ($I_{c,\mathrm{'1'}}$) $I_c$ values, respectively.  Therefore, with a suitable current ($I_{c,\mathrm{'1'}} < I_{DS} < I_{c,\mathrm{'0'}}$), the Fe--JoFET channel will show a zero voltage for $P_R^-$ and non-zero voltage for $P_R^+$ states. 

\subsection{Validation}
Our proposed model is validated against experimental data from Ref. \cite{Paghi2025FeJoFET}. Initially, the values of $P_S$, $P_R$, and $V_c$ are extracted. Then, using the Preisach block discussed in Section~\ref{Methodology}, our model accurately portrays the $P-V$ characteristics of the experimental device (Fig. \ref{fig:FeJoFET_Characteristics_Exploration}(a)). Next, we reproduce the effect of $V_{GS}$ on $I_c$ using our modified AB theory, mimicking the experimental behavior (Fig. \ref{fig:FeJoFET_Characteristics_Exploration}(b)). We also model the effect of $V_{GS}$ on $R_N$ using a look-up table-based approach (Fig. \ref{fig:FeJoFET_Characteristics_Exploration}(d)). Finally, using the RCSJ model, we demonstrate the validated $I-V$ characteristics of the Fe--JoFET at zero gate voltage in Fig. \ref{fig:FeJoFET_Characteristics_Exploration}(e). Thus, our proposed compact model is able to successfully replicate the device behavior as reported in Ref. \cite{Paghi2025FeJoFET} where all measurements are taken from a Fe--JoFET with $5~\mu{m}$ width and $600~{nm}$ length at an operating temperature of $50~{mK}$. Furthermore, the close agreement between the experiments and the simulation indicates that our model is suitable for circuit-level use.

\subsection{Design Space Exploration}
After validation, we check the ability of our model to reproduce device behaviors in different combinations of ferroelectric $P$ and $V_{GS}$. Fig. \ref{fig:FeJoFET_Characteristics_Exploration}(c) and (f) show the changes of $I_c$, $R_N$, and $V_{DS}$ for various $V_{GS}$ values at different polarization states. The primary objective is to observe the device's response and facilitate device engineering for different potential applications. Fig. \ref{fig:FeJoFET_Characteristics_Exploration}(c) shows that $I_c$ and $R_N$ vary strongly with $V_{GS}$, but with opposite trends. Fig. \ref{fig:FeJoFET_Characteristics_Exploration}(f) represents $V_{DS}$ as a function of $V_{GS}$ for a fixed $I_{DS} = 300 nA$. Both states share the same SC and resistive regions, but are differentiated by their distinct $I_c$ values. From these curves, we can select a suitable $V_{GS}$ to obtain a specific combination of $I_c$ and $R_N$. Therefore, our model can predict how Fe--JoFETs can be programmed in suitable conditions for potential circuit/system-level applications. For example, we use these insights to choose suitable read and write biases in our proposed memory array, as explained in Sections IV and V.

\section{Fe--JoFET-based Memory Array}
\subsection{Design Principle}

\begin{figure*}[t]
    \centering
    \includegraphics[width=0.6\textwidth]{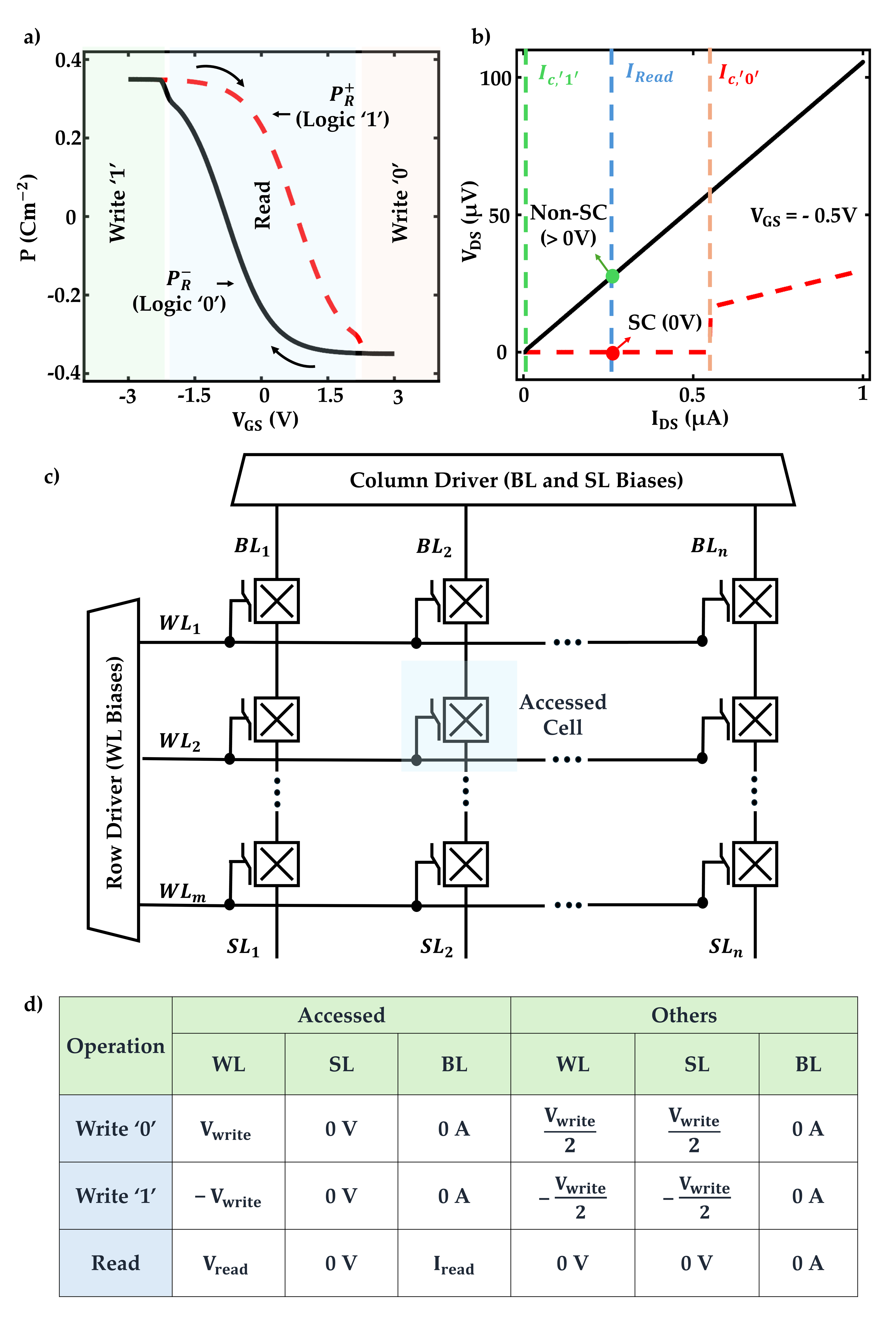}
    \caption{Fe--JoFET-based cryogenic memory array. \textbf{(a)} $P$-$V$ characteristics divided into write and read regions. \textbf{(b)} $V_{\mathrm{DS}}$ vs. $I_{\mathrm{DS}}$, illustrating the state-dependent $I_{\mathrm{c,'1'}}$ and $I_{\mathrm{c,'0'}}$. The read current $I_{\mathrm{read}}$ is selected between them so that the $P_{\mathrm{R}}^{-}$ state remains SC, whereas the $P_{\mathrm{R}}^{+}$ state becomes resistive. \textbf{(c)} Schematic of the proposed selectorless $m\times n$ memory array. One randomly selected cell is highlighted to showcase the write and read operations on that particular cell. \textbf{(d)} Bias conditions for accessing the selected cell during write '0', write '1', and read operations.}
    \label{fig:FeJoFET_Memory_Array}
\end{figure*} 

We utilize the two distinct polarization states of the ferroelectric material to define binary logic states. The $P_R^-$ and $P_R^+$ states correspond to logic '0' and logic '1', respectively. Thus, stored logic '0' will have a higher $I_c$ and lower $R_N$ than that of a stored logic '1' cell. This difference in $I_c$ allows the Fe--JoFET to act as an individually readable memory cell. Fig.~\ref{fig:FeJoFET_Memory_Array}(a) shows the $P$--$V_{GS}$ characteristics of the device divided into three regions: (i) Write '0', (ii) Read, and (iii) Write '1', highlighting the $V_{GS}$ region applicable for each operation. A substantially positive (negative) $V_{GS}$ is required to store logic '0' (logic '1') in the memory cell, whereas the read operation can occur with $V_{GS}$ between these switching voltages. 

Fig.~\ref{fig:FeJoFET_Memory_Array}(b) shows an example of the read operation of an individual cell at a particular $V_{GS}$. The two curves highlight two distinct $I_c$ values of $I_{c,'0'}$ and $I_{c,'1'}$ directly corresponding to its ferroelectric polarization states of $P_R^-$ and $P_R^+$, respectively. By choosing a read current between $I_{c,'1'}$ and $I_{c,'0'}$, the $P_R^-$ state outputs an SC state (zero voltage), whereas the $P_R^+$ state outputs a resistive state (non-zero voltage). Thus, the stored logic can be determined from the $V_{DS}$ output of a particular Fe--JoFET memory cell with a suitable $I_{DS}$.

Fig.~\ref{fig:FeJoFET_Memory_Array}(c) shows the proposed $m\times n$ memory array. The gates of the cells in each row
are connected to a common word line ($\mathrm{WL}$), while the drain terminal of the first cell in each column is connected to a bit line ($\mathrm{BL}$) and the source terminal of the last cell in each column is connected to a source line ($\mathrm{SL}$). The source and drain terminals of other cells in a column are connected in a way so that all the cells in a column are connected in series. Therefore, a cell is selected by applying a suitable combination of $\mathrm{WL}$, $\mathrm{BL}$, and $\mathrm{SL}$ biases. Unlike conventional SC memory arrays, our proposed structure does not require an additional selector to access a particular cell. Furthermore, as seen in Fig.~\ref{fig:FeJoFET_Memory_Array}(b), a suitable read current can lead to zero and non-zero voltages for logic '0' and '1', respectively. Thus, the stored information of a particular cell can be ascertained from the $\mathrm{BL}$ voltage without any sensing circuitry. 

\begin{figure*}[t]
    \centering
    \includegraphics[width=1\textwidth]{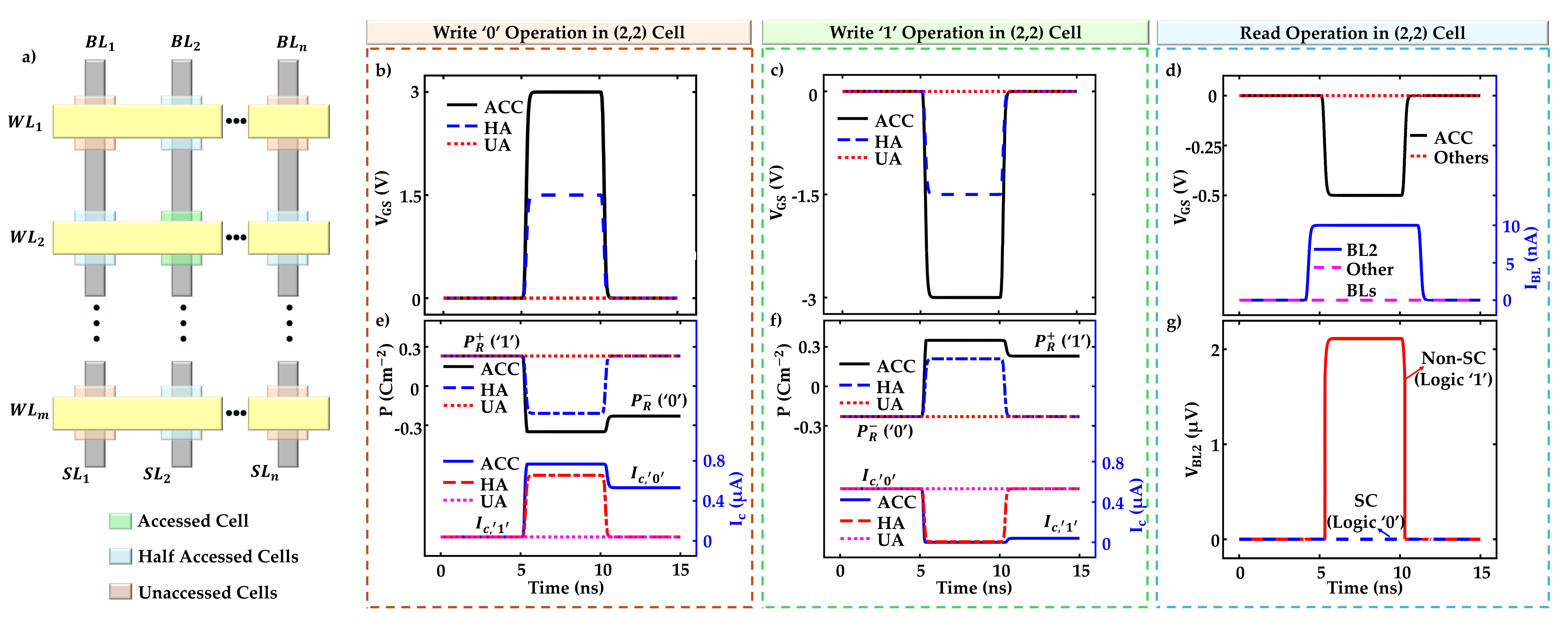}
    \caption{Simulation of write and read operations in a specific cell in the array. \textbf{(a)} Proposed $m\times n$ array organization showing the accessed (ACC), half-accessed (HA), and unaccessed (UA) cells when accessing the cell highlighted in Fig.~\ref{fig:FeJoFET_Memory_Array}(c). $V_{GS}$ and $I_{BL}$ of the cells during \textbf{(b)} write '0', \textbf{(c)} write '1', and \textbf{(d)} read operations. Corresponding $P$ and $I_c$ transitions during \textbf{(e)} write '0' and \textbf{(f)} write '1' operations. \textbf{(g)} The resulting bit-line voltage, which is zero for logic '0' state and nonzero for logic '1' state.}
    \label{fig:FeJoFET_Write_Read}
\end{figure*} 

\subsection{Working Principle}
Fig.~\ref{fig:FeJoFET_Memory_Array}(d) shows the biasing scheme of our Fe--JoFET-based memory array for both write and read operations. Without any loss of generality, here we choose an arbitrary memory cell (2,2) as highlighted in Fig.~\ref{fig:FeJoFET_Memory_Array}(c) to demonstrate the memory operations. For write operations, we utilize a $V/2$ biasing scheme where $\mathrm{WL}$ and $\mathrm{SL}$ of the accessed cell are biased at an appropriate write voltage ($\pm V_{write}$) and $0V$, respectively. All other $\mathrm{WLs}$ and $\mathrm{SLs}$ are biased at $\pm V_{write}/2$ voltage.  At the same time, all $\mathrm{BLs}$ are biased with zero current throughout the writing process, just to ensure that all the cells in the array remain SC for any combinations of polarization state and $V_{GS}$. $V_{write}$ is chosen to be larger than the absolute switching voltage ($V_{switch}$) at which the device's polarization state changes, but $V_{write}/2$ is less than $|V_{switch}|$. This is done so that only the accessed cell gets sufficient voltage across it. In short, the conditions for choosing $V_{write}$ are- 
\begin{equation}
V_{\mathrm{write}} > |V_{\mathrm{switch}}| \  \&\& \  
\frac{V_{\mathrm{write}}}{2} < |V_{\mathrm{switch}}| 
\end{equation} 

For read operations, the $\mathrm{WL}$ connected to the accessed cell is biased at a suitable read voltage ($V_{read}$), while the $\mathrm{BL}$ and $\mathrm{SL}$ connected to the accessed cell are supplied with a suitable read current ($I_{read}$) and $0V$, respectively. During read, other $\mathrm{WLs}$ and $\mathrm{SLs}$ are grounded, while other $\mathrm{BLs}$ are supplied with no current. The conditions of choosing $V_{read}$ and $I_{read}$ are-

\begin{equation}
\begin{aligned}
\left|V_{\mathrm{read}}\right|
&< \left|V_{\mathrm{switch}}\right| \&\&\\
I_{\mathrm{c}}\!\left(P=P_{\mathrm{R}}^{+},
V_{\mathrm{GS}}=V_{\mathrm{read}}\right)
&<
I_{\mathrm{c}}\!\left(P=P_{\mathrm{R}}^{+},
V_{\mathrm{GS}}=0~\mathrm{V}\right)
\end{aligned}
\label{eq:read_voltage_condition}
\end{equation}

\begin{equation}
\begin{aligned}
I_{\mathrm{c}}\!\left(P=P_{\mathrm{R}}^{+},
V_{\mathrm{GS}}=V_{\mathrm{read}}\right)
&< I_{\mathrm{read}} \\
&< \min\!\left(
\begin{gathered}
I_{\mathrm{c}}\!\left(P=P_{\mathrm{R}}^{+},
V_{\mathrm{GS}}=0~\mathrm{V}\right),\\
I_{\mathrm{c}}\!\left(P=P_{\mathrm{R}}^{-},
V_{\mathrm{GS}}=V_{\mathrm{read}}\right)
\end{gathered}
\right)
\end{aligned}
\label{eq:read_current_condition}
\end{equation}

This combination of voltage and current biases ensure that only the accessed cell gets the required combination for the read operation. Another requirement is that all other cells except the accessed cell must remain SC during the read operation. For the chosen $I_{read}$, the $P_R^-$ ($P_R^+$) state corresponds to the SC (resistive) state of the Fe-JoFET channel. Since all the other cells remain SC, the voltage drop across the accessed cell will determine the $\mathrm{BL}$ voltage which can be used to identify the stored memory state. A zero voltage means logic '0', whereas a nonzero voltage means logic '1', directly providing a binary decision without any additional peripheral circuits. 

\section{Memory Operation}
Using the developed compact model (discussed in Section IIIA), we simulate the write and read operations in HSPICE to verify the functionality of the proposed memory array. Fig.~\ref{fig:FeJoFET_Write_Read} shows the simulated write and read operations for the (2,2) cell in the array. Fig.~\ref{fig:FeJoFET_Write_Read}(a) illustrates the division of the array into three types of cells during memory operations: (i) accessed cell (ACC), (ii) half accessed cells (HA) and (iii) unaccessed cells (UA). All the cells in the same row and column as the ACC cell are the HA cells, and all other cells are the UA cells. The ACC cell gets the full $\pm V_{write}$ across its gate-source, whereas the HA cells get $\pm V_{write}/2$, and finally the UA cells get 0 V, as shown in Fig.~\ref{fig:FeJoFET_Write_Read}(b) and (c). All the BLs have zero current, ensuring all the cells remain SC. This allows the necessary $V_{GS}$ values for every cell so that only the ACC cell changes its state and all the other cells remain unchanged. Fig.~\ref{fig:FeJoFET_Write_Read}(e) and (f) show that only the ACC cell has its polarization state and $I_c$ value permanently changed from $P_R^+$ ($P_R^-$) to $P_R^-$ ($P_R^+$) and $I_{c,'1'}$ ($I_{c,'0'}$) to $I_{c,'0'}$ ($I_{c,'1'}$) during the write '0' (write '1') operation. Every other cell remains in its previous state upon the removal of biases. Thus, write operations can be performed to any specific cell even without a selector. The write operations of Fig.~\ref{fig:FeJoFET_Write_Read} use $V_{write} = 3 V$. 

Fig.~\ref{fig:FeJoFET_Write_Read}(d) shows the $V_{GS}$ and $I_{BL}$ values for the read operation of the accessed cell. During read, only the ACC cell gets $V_{read}$ across its gate and source, with all the other cells getting 0 V. Similarly, only the BL of the ACC cell is biased with $I_{read}$, while every other $\mathrm{BL}$ is biased with 0 A. Because all cells except the ACC cell are SC during read, the drain-source voltage of the accessed cell will be the BL voltage. Therefore, the BL voltage will be zero or non-zero depending on the memory state, as illustrated in Fig.~\ref{fig:FeJoFET_Write_Read}(g). The read operation demonstrated in Fig.~\ref{fig:FeJoFET_Write_Read} uses $V_{read} = -0.5 V$ and $I_{read} = 10 nA$.

\section{Benchmark}

\begin{table*}[t]
\centering

\caption{
Comparison of the proposed selectorless Fe--JoFET memory array
with existing cryogenic memory technologies.
}

\label{tab:memory_comparison}

\scriptsize
\setlength{\tabcolsep}{3pt}
\renewcommand{\arraystretch}{1.25}

\begin{threeparttable}

\begin{tabularx}
{\textwidth}
{@{}L{2.25cm}*{8}{Y}@{}}

\toprule

\makecell[c]{
  \textbf{Memory}
}
&
\makecell[c]{
  \textbf{Volatility}
}
&
\makecell[c]{
  \textbf{Selectorless}\\
  \textbf{Cell?}
}
&
\makecell[c]{
  \textbf{Cell}
  \textbf{Area}\\
  \textbf{($\mu\mathrm{m}^{2}$)}
}
&
\makecell[c]{
  \textbf{Power/Cell}
}
&
\makecell[c]{
  \textbf{Inductorless}\\
  \textbf{Cell?}
}
&
\makecell[c]{
  \textbf{Magnetic-Bias}\\
  \textbf{Required?}
}
&
\makecell[c]{
  \textbf{Sensing }\\
  \textbf{Circuit?}
}
&
\makecell[c]{
  \textbf{Operating}\\
  \textbf{Temperature}
}
\\

\midrule

Cryo-CMOS/eDRAM
\cite{Shu2023CSDBeDRAM}
&
\textcolor{red}{Volatile}
&
\textcolor{red}{No}
&
\textcolor{green}{$\sim0.557$}
&
$49.23~\mu\mathrm{W}$
&
\textcolor{green}{Yes}
&
\textcolor{green}{No}
&
\textcolor{red}{Yes}
&
$4.2~\mathrm{K}$
\\
\midrule
SFQ CRAM
\cite{Kirichenko1999}
&
\textcolor{red}{Volatile}
&
\textcolor{red}{No}
&
$40\times45$
&
  $\sim0.15~\mu\mathrm{W}$
&
\textcolor{red}{No}
&
\textcolor{green}{No}
&
\textcolor{red}{Yes}
&
4 K
\\
\midrule
Magnetic JJ Memory
\cite{Ryazanov2012MJJMemory}
&
\textcolor{green}{nonvolatile}
&
\textcolor{red}{No}
&

  $\sim16$
  $MJJ \ Contact$
&
Not Reported
&
\textcolor{green}{Yes}
&
\textcolor{red}{Yes}
&
\textcolor{red}{Yes}
&
$4.2~\mathrm{K}$
\\
\midrule
Superconducting Memristor Array
\cite{Alam2021CryoMemristor}
&
\textcolor{green}{nonvolatile}
&
\textcolor{red}{No}
&
  Not Reported
&

Read: $1.15~\mu\mathrm{W}$;
Write: $10.1~\mu\mathrm{W}$
(Estimated)

&
\textcolor{green}{Yes}
&
\textcolor{red}{Yes}
&
\textcolor{red}{Yes}
&
$1~\mathrm{K}$
\\

\midrule
FeSQUID Memory Array
\cite{Alam2022FeSQUIDMemory, Suleiman2021FeSQUIDMemory}
&
\textcolor{green}{nonvolatile}
&
\textcolor{red}{No}
&
Not Reported
&
Read: $30~\mathrm{nW}$;
Write: No power
&
\textcolor{green}{Yes}
&
\textcolor{green}{No}
&
\textcolor{green}{No}
&
$\leq4~\mathrm{K}$
\\
\midrule
Hybrid Josephson--CMOS Memory
\cite{Liu2007HybridJCMOS}
&
\textcolor{red}{Volatile}
&
\textcolor{red}{No}
&
$6.5\times7.5$
&
Read: $0.01~\mu\mathrm{W}$;
Write: $0.02~\mu\mathrm{W}$
&
\textcolor{green}{Yes}
&
\textcolor{green}{No}
&
\textcolor{red}{Yes}
&
$4.2~\mathrm{K}$
\\

\midrule

\textbf{This work}
&
\textcolor{green}{nonvolatile}
&
\textcolor{green}{Yes}
&
$\sim3$
&
\textcolor{green}{Read: $10.55~\mathrm{fW}$;
Write: No power}
&
\textcolor{green}{Yes}
&
\textcolor{green}{No}
&
\textcolor{green}{No}
&
$50~\mathrm{mK}$
\\

\bottomrule

\end{tabularx}

\end{threeparttable}

\end{table*}

Table~\ref{tab:memory_comparison} compares the proposed
selectorless Fe--JoFET memory array with representative
cryogenic memory technologies. Cryogenic memories must provide
high density, fast operation, and nonvolatile storage while
operating within the limited cooling power available at low
temperatures
\cite{Alam2023CryogenicMemory,Hornibrook2015CryogenicControl,
Tannu2017CryoDRAM}. Cryo-CMOS (such as SRAM and DRAM) and other non-SC memories provide technological maturity and high storage capacity, but remain affected by volatility, refresh overhead, and conventional sensing circuitry \cite{Tannu2017CryoDRAM,Alam2023CryogenicMemory}. Their high power consumption is also a major challenge for cryogenic environments. Conventional SC memories,
such as SFQ CRAM, magnetic JJ-based, and SC memristor-based memories provide high speed and consume extremely low power. However, they generally require SQUID loops, inductive or magnetic coupling, current-driven select lines,
sensing gates, and substantial peripheral circuitry
\cite{Likharev1991RSFQ,Holmes2013EnergyBudget,
Tolpygo2016Scalability,Kirichenko1999}. The SFQ CRAM and magnetic JJ-based memories
illustrate these limitations through their large cell area. Therefore, they suffer from extremely poor scalability. On the other hand, hybrid
Josephson--CMOS memories combine the advantages of non-SC and SC technologies, but suffer from the lack of a suitable interface circuit \cite{Liu2007HybridJCMOS, Alam2023CryogenicMemory}. 

The SC memristor array
provides nonvolatile resistance states, but requires an external magnetic-flux bias, a heater cryotron \cite{alam2023cryogenic} access device, and cryogenic sensing circuitry \cite{Alam2021CryoMemristor}. The FeSQUID memory removes the need for magnetic bias, external inductive coupling, and bulky peripheral sensing circuitry, but
still relies on a heater cryotron for cell access
\cite{Suleiman2021FeSQUIDMemory,Alam2022FeSQUIDMemory}.
Therefore, as summarized in
Table~\ref{tab:memory_comparison}, existing nonvolatile
approaches improve energy efficiency but need additional
selectors, magnetic biasing, inductive elements, or sensing
circuits that increase array complexity and limit
scalability.

The proposed nonvolatile Fe--JoFET array addresses these limitations by
using the remnant polarization of the ferroelectric gate to
control the operating state of the SC channel. The ferroelectric gate of Fe--JoFET provides both
storage and intrinsic cell selection, eliminating the need
for an additional selector device. During read operation, the stored
state is determined directly from the bit-line voltage,
eliminating a dedicated cryogenic sensing circuit. As a
result, the proposed architecture avoids external inductive
coupling, magnetic biasing, selector dissipation, and
dedicated sensing power. Therefore, its cell-level operating power is lower than any of the existing technologies. In addition, the cell area is also smaller than any existing SC or hybrid technologies because of not needing any additional access device. 

\section{Conclusion}

In this work, we present a highly scalable, selectorless, and ultra-low power cryogenic memory based on Fe--JoFET. First, we develop a physics-based Verilog-A compact model for this device to facilitate its circuit and system-level exploration. We use this experiment-calibrated model to verify the functionality of the proposed memory array. The remnant
ferroelectric polarization provides nonvolatile storage, while
the Fe--JoFET itself provides intrinsic cell selection. Because
all unselected cells remain superconducting during read and
write operations, the proposed architecture avoids additional
selectors, external inductors, magnetic biasing, and dedicated
sensing circuits. This device-to-array co-design therefore
offers a promising approach to reducing peripheral complexity
and operating power while improving the scalability of
superconducting cryogenic memories. The proposed architecture
can be useful in quantum computing, SC computing,
and other energy-constrained systems. Additionally, it can potentially revolutionize the field of cryogenic neuromorphic systems due to its very low power consumption while retaining the speed of ferroelectric switching. Thus, future work focusing on retention and disturb analysis, and experimental validation of the complete memory architecture is imperative. 

\section{References}
\vspace{-3.5ex}
\bibliography{references}
\end{document}